\documentclass[11pt,letterpaper]{article}
\usepackage{osajnl2}
\usepackage{xspace}
\usepackage{graphicx}
\newcommand{\ignore}[1]{}
\renewcommand{\ao}{adaptive optics (AO)\renewcommand{\ao}{AO\xspace}\xspace}
\newcommand{\wfs}{wavefront sensor (WFS)\renewcommand{\wfs}{WFS\xspace}\renewcommand{\wfss}{WFSs\xspace}\xspace}
\newcommand{\wfss}{wavefront sensors (WFSs)\renewcommand{\wfs}{WFS\xspace}\renewcommand{\wfss}{WFSs\xspace}\xspace}
\newcommand{\dm}{deformable mirror (DM)\renewcommand{\dm}{DM\xspace}\renewcommand{\dms}{DMs\xspace}\xspace}
\newcommand{\dms}{deformable mirrors (DMs)\renewcommand{\dm}{DM\xspace}\renewcommand{\dms}{DMs\xspace}\xspace}
\newcommand{\shs}{Shack-Hartmann sensor (SHS)\renewcommand{\shs}{SHS\xspace}\renewcommand{\shss}{SHSs\xspace}\xspace}
\newcommand{\shss}{Shack-Hartmann sensors (SHSs)\renewcommand{\shs}{SHS\xspace}\renewcommand{\shss}{SHSs\xspace}\xspace}
\newcommand{\lgs}{laser guide star (LGS)\renewcommand{\lgs}{LGS\xspace}\xspace}
\newcommand{\ngs}{natural guide star (NGS)\renewcommand{\ngs}{NGS\xspace}\xspace}
\newcommand{\mems}{Micro-Electro-Mechanical Systems (MEMS)\renewcommand{\mems}{MEMS\xspace}\xspace}
\newcommand{\snr}{signal to noise ratio (SNR)\renewcommand{\snr}{SNR\xspace}\xspace}
\newcommand{\moao}{multi-object adaptive optics (MOAO)\renewcommand{\moao}{MOAO\xspace}\xspace}
\newcommand{\ltao}{laser tomographic adaptive optics (LTAO)\renewcommand{\ltao}{LTAO\xspace}\xspace}
\newcommand{\cpu}{central processing unit (CPU)\renewcommand{\cpu}{CPU\xspace}\xspace}
\newcommand{\psf}{point spread function (PSF)\renewcommand{\psf}{PSF\xspace}\renewcommand{\psfs}{PSFs\xspace}\xspace}
\newcommand{\psfs}{point spread functions (PSFs)\renewcommand{\psf}{PSF\xspace}\renewcommand{\psfs}{PSFs\xspace}\xspace}
\newcommand{\fpga}{field programmable gate array (FPGA)\renewcommand{\fpga}{FPGA\xspace}\renewcommand{\fpgas}{FPGAs\xspace}\xspace}
\newcommand{\fpgas}{field programmable gate arrays (FPGAs)\renewcommand{\fpga}{FPGA\xspace}\renewcommand{\fpgas}{FPGAs\xspace}\xspace}
\newcommand{\sor}{successive over-relaxation (SOR)\renewcommand{\sor}{SOR\xspace}\xspace}
\newcommand{\fdpcg}{Fourier domain pre-conditioned gradient (FDPCG)\renewcommand{\fdpcg}{FDPCG\xspace}\xspace}
\newcommand{\map}{maximum a-posteriori (MAP)\renewcommand{\map}{MAP\xspace}\xspace}
\newcommand{\elt}{extremely large telescope (ELT)\renewcommand{\elt}{ELT\xspace}\xspace}
\newcommand{\dugall}{Durham University generalised adaptive optics laser laboratory (DUGALL)\renewcommand{\dugall}{DUGALL\xspace}\xspace}
\newcommand{\fwhm}{full-width at half-maximum (FWHM)\renewcommand{\fwhm}{FWHM\xspace}\xspace}
\newcommand{\wht}{William Herschel Telescope (WHT)\renewcommand{\wht}{WHT\xspace}\xspace}
\newcommand{\emccd}{electron multiplying CCD (EMCCD)\renewcommand{\emccd}{EMCCD\xspace}\xspace}
\newcommand{\obshs}{Optically binned SHS (OBSHS)\renewcommand{\obshs}{OBSHS\xspace}\xspace}

\newcommand{\mnras}{MNRAS}
\newcommand{\citep}[1]{\cite{#1}}
\begin{document}
\title{Shack-Hartmann sensor improvement using optical binning}
\author{Alastair Basden, Deli Geng, Dani Guzman, Tim Morris, Richard
  Myers, Chris Saunter}
\address{Department of Physics, South Road, Durham, DH1 3LE, UK}

\begin{abstract}
We present a design improvement for a recently proposed type of
Shack-Hartmann wavefront sensor that uses a cylindrical (lenticular)
lenslet array.  The improved sensor design uses optical binning and
requires significantly fewer detector pixels than the corresponding
conventional or cylindrical Shack-Hartmann sensor, and so detector
readout noise causes less signal degradation.  Additionally, detector
readout time is significantly reduced, which reduces the latency for
closed loop systems, and data processing requirements.  We provide
simple analytical noise considerations and Monte-Carlo simulations,
and show that the optically binned Shack-Hartmann sensor can offer
better performance than the conventional counterpart in most practical
situations, and our design is particularly suited for use with
astronomical adaptive optics systems.
\end{abstract}

\maketitle

\section{Introduction}
\begin{figure}
\includegraphics[width=5cm]{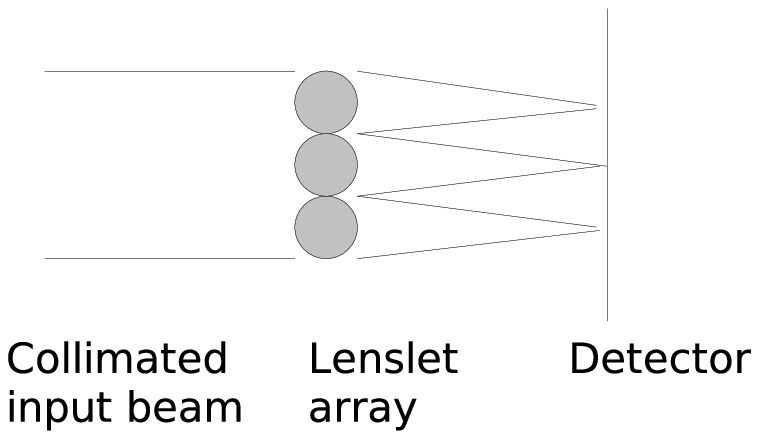}\includegraphics[width=3cm]{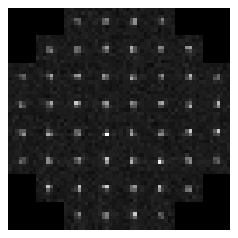}
\caption{A schematic diagram of a conventional Shack-Hartmann
  wavefront sensor, and a typical spot pattern with 64 sub-apertures
  (4096 detector pixels).}
\label{fig:conventional}
\end{figure}

A conventional \shs divides a pupil into sub-apertures using a lenslet
array and attempts to measure the wavefront gradients in orthogonal
directions across each sub-aperture as shown in
Fig.~\ref{fig:conventional}.  Estimation of the wavefront gradients
typically involves finding the centre-of-mass of the image spot
created in the sub-aperture (the mean light position).  This is
typically done by software binning of the measured light signal in one
direction when computing the algorithm, and then computing the dot
product of this vector with an index vector (i.e.\ a vector counting
from 0 to N-1 where N is the number of pixels in a sub-aperture).
Computation of the corresponding orthogonal wavefront tilt is carried
out by software binning the measured light signal in the orthogonal
direction.  Once these spot centroid locations have been retrieved, a
reconstruction algorithm is used to provide an estimate of the
wavefront under investigation.

Here, we present a modified version of the \shs described by
Ares et al.\cite{cylindricalSH}.  This design uses the cylindrical lenslet array
proposed previous, and also implements optical signal binning.  There
are a number of situations where this can give a performance
improvement when compared with a conventional \shs which we describe
in \S\ref{sect:design}.  The original cylindrical Shack-Hartmann
sensor design described by Ares\cite{cylindricalSH} was intended to
provide a reliable way to measure wavefront gradients outside the
nominal Shack-Hartmann lenslet area on the detector, such as when
there are highly aberrated wavefronts or abrupt phase changes, for
time-static aberrations.  The design proposed here has an additional
aim, to achieve higher \wfs frame-rates than would be possible using a
conventional \shs, by using fewer detector pixels, and also to give
improved \snr performance.

\begin{figure}
(a)\\
\includegraphics[width=8cm]{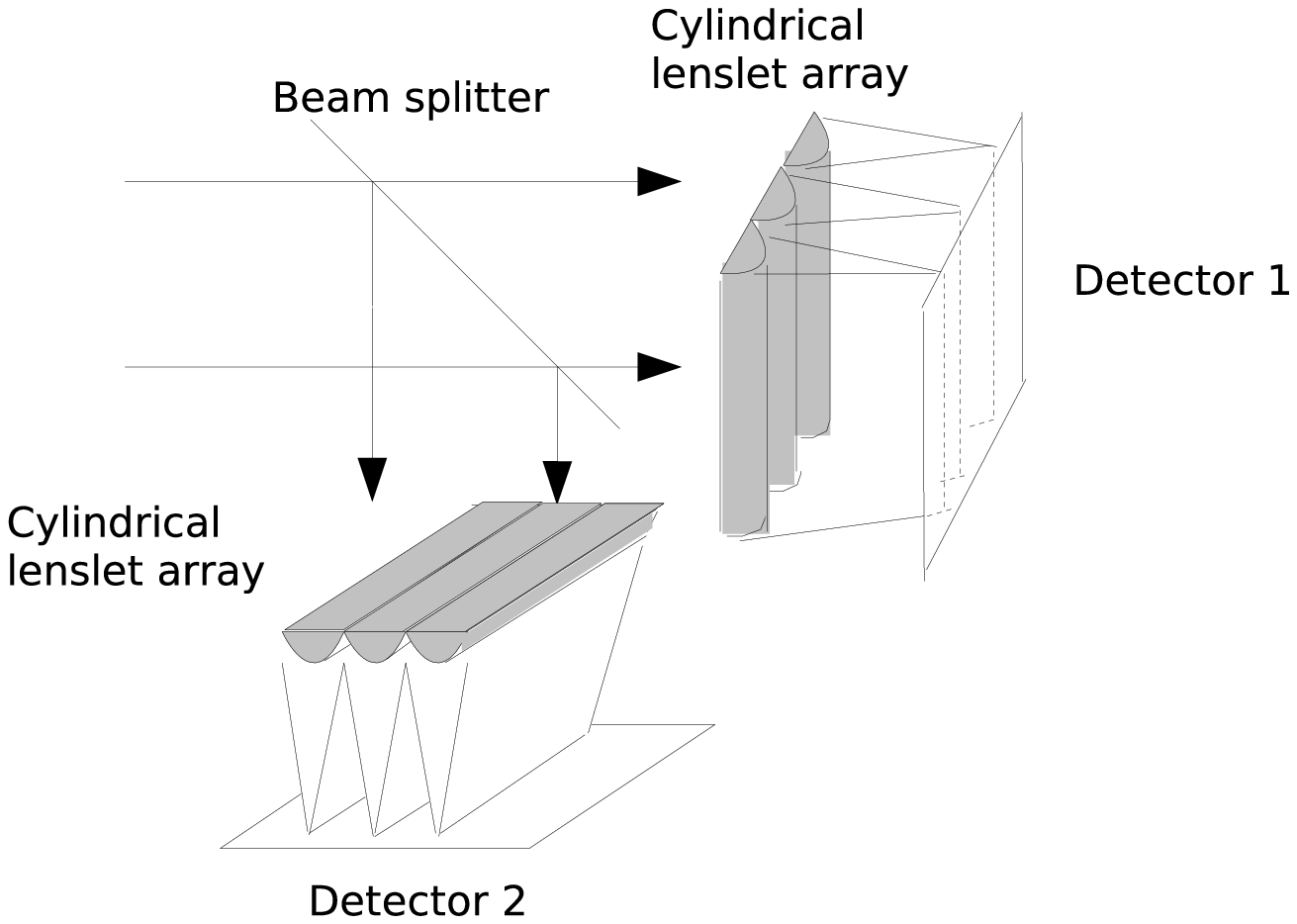}\\
(b)\hspace{4cm}(c)\\
\includegraphics[width=4cm,height=4cm]{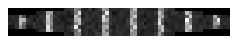}
\includegraphics[angle=90,width=4cm,height=4cm]{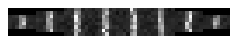}
\caption{(a) A schematic diagram of an optically binned Shack-Hartmann
  wavefront sensor.  The incoming wavefront is split using a beam
  splitter, and each beam then passes through orthogonal cylindrical
  lenslet arrays, to record the x and y wavefront gradients on
  separate detectors.  (b) A typical resulting image from detector 1
  (512 detector pixels, $8\times8$ sub-apertures, vertical lenslet
  array).  (c) A typical resulting image from detector 2 (512 detector
  pixels, $8\times8$ sub-apertures, horizontal lenslet array).
  Elongated rectangular detector pixels have been used in (b) and (c)
  to make the image clearer.}
\label{fig:opticalbin}
\end{figure}

\begin{figure}
\includegraphics[width=8cm]{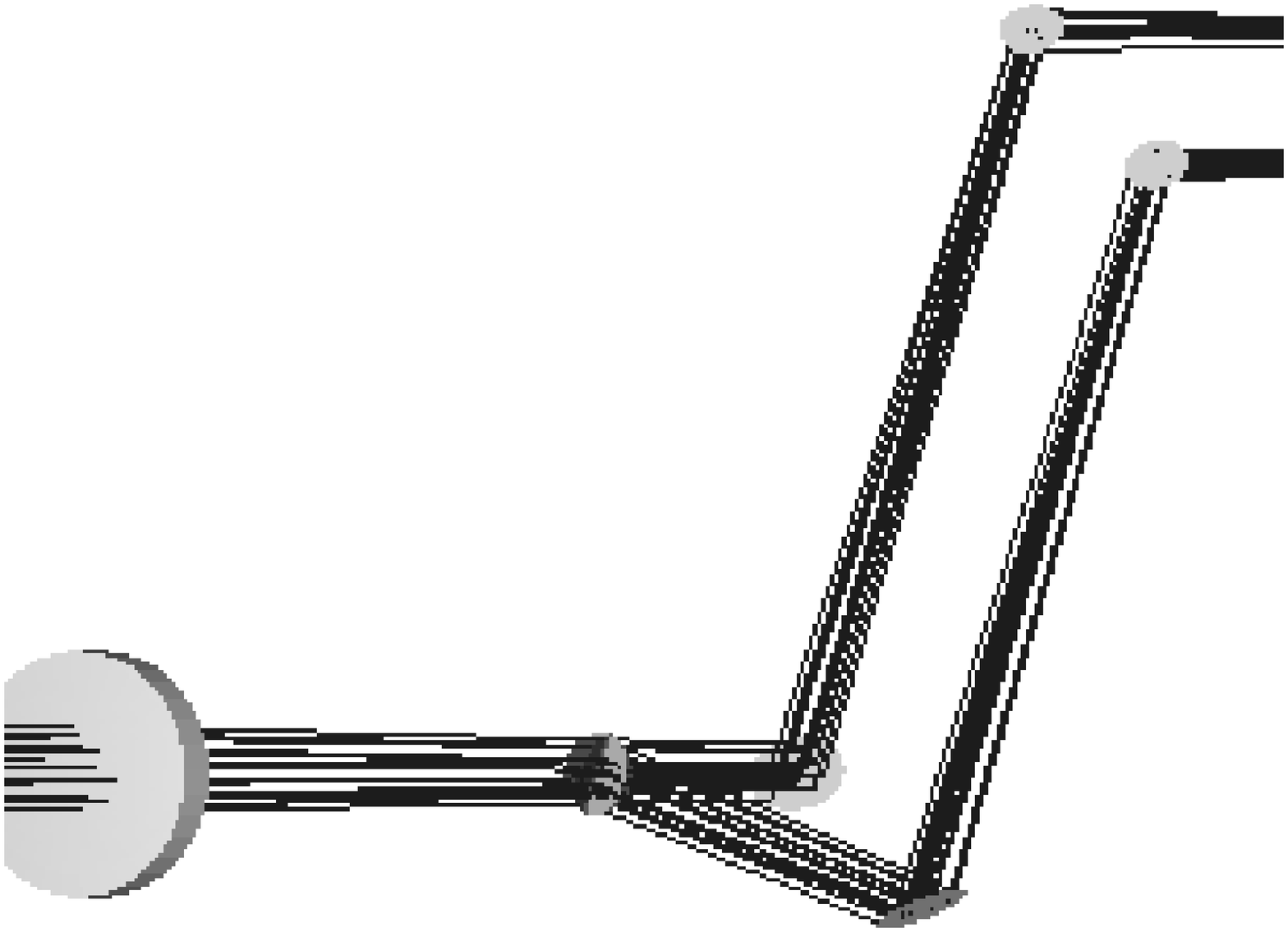}
\includegraphics[width=8cm]{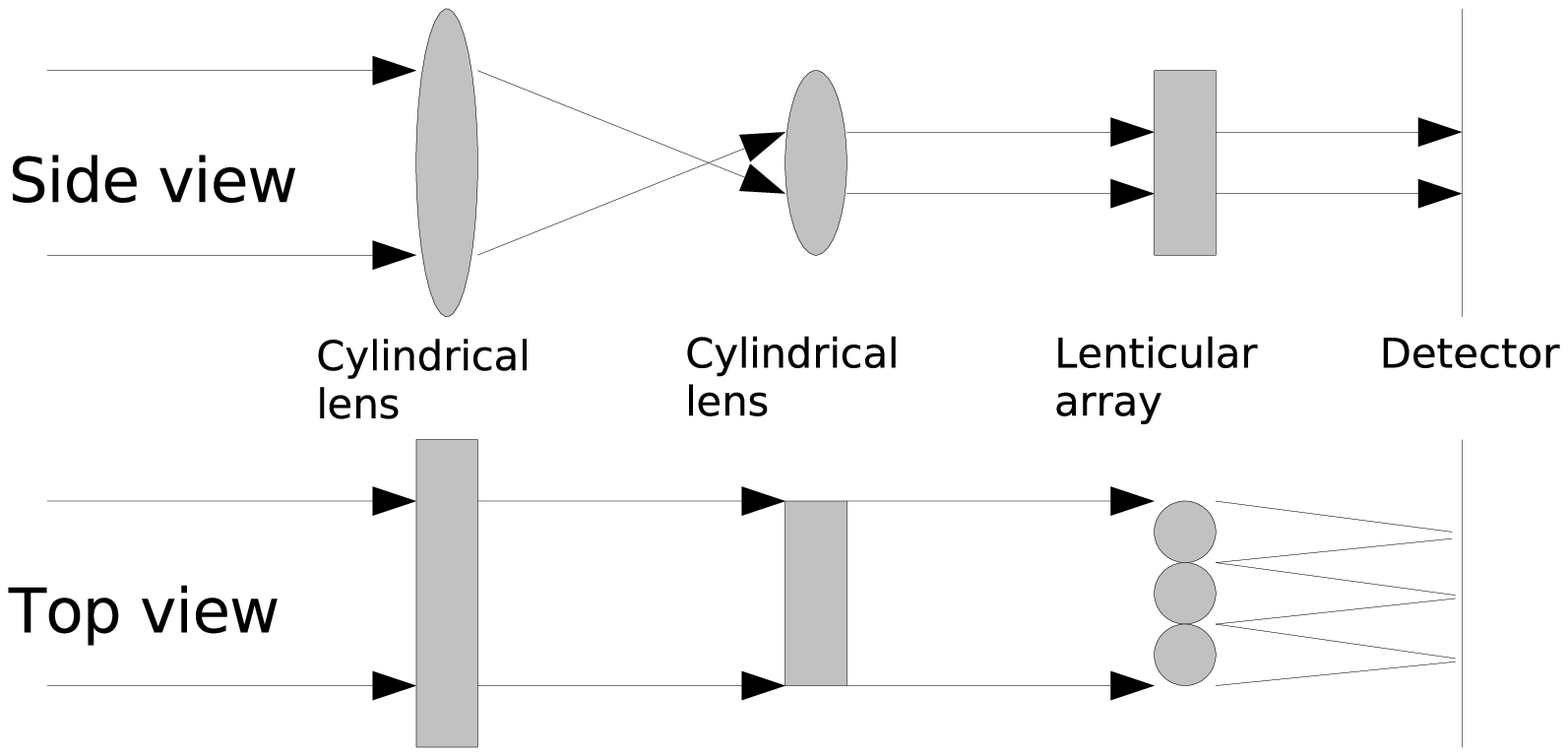}
\caption{A schematic diagram showing possible sub-designs for an
  optically binned Shack-Hartmann sensor, (a) using fold mirrors to
  rotate the reflected beam by $90^\circ$ (b) using shared cylindrical
  optics for each beam, one beam shown, to compress the phase in one
  dimension relative to the other.}
\label{fig:design}
\end{figure}

A simplified schematic of an \obshs is given in
Fig.~\ref{fig:opticalbin}, along with an example for the detector
images.  The wavefront is first split with a 50/50 beam-splitter.
Each of the resulting beams then passes through one of two identical
cylindrical (lenticular) lenslet arrays oriented orthogonally, some
cylindrical re-imaging optics (not shown here for clarity, see
Fig.~\ref{fig:design}), and onto separate detectors.  A conventional
(circular) lenslet array focuses the wavefront in two directions,
giving a conventional point-spread function.  Conversely, the
cylindrical lenslet arrays proposed here focus the wavefront in only
one direction.  Rather than a single spot, light will be spread along
a number of lines as shown in Fig.~\ref{fig:opticalbin}.

The two orthogonal lenslet arrays are required to measure orthogonal
wavefront gradients.  The design described by Ares\cite{cylindricalSH}
used a single cylindrical array which was rotated by $90^\circ$ to
measure the orthogonal wavefront gradients, rather than the beam
splitter shown here.  A drawback of this technique is that it is
only suitable for characterising static aberrations, as detector
images have to be captured before and after the precise $90^\circ$
rotation.

In order to minimise the number of detector pixels required for
wavefront gradient estimation, the \obshs should be designed such that
the width of the focused line (Fig.~\ref{fig:opticalbin}(b) and (c))
is $n_s$ detector pixels wide where $n_s$ is the number of
sub-apertures in each dimension (i.e.\ there are $n_s\times n_s$
sub-apertures in total).  This means that a one-dimensional
centre-of-mass calculation for the line position in the direction
orthogonal to the line will give the corresponding centroid location
and wavefront gradient estimate in this direction for this
sub-aperture, i.e.\ by measuring the offset of the line from the
nominal origin position of each sub-aperture.  Each sub-aperture is
one pixel wide and a larger number ($n_l$, e.g.\ eight) pixels long,
depending on the field of view required.  When the advanced processing
techniques for line location are used \citep{cylindricalSH}, the
number of pixels orthogonal to the cylindrical lenslet direction
($n_l$) can be reduced compared with a conventional \shs, whilst
achieving the same field of view.  These processing techniques
effectively apply a continuity condition to the measured line
position, and greatly reduce the problem of sub-aperture cross-talk,
except for when the local wavefront tilt is great enough to place
light from one sub-aperture on top of another.

Unless the detector has elongated pixels, it will be necessary to use
non-symmetrical re-imaging optics to compress the image in one
dimension, such that each sub-aperture is then one pixel wide and
$n_l$ pixels long (so that the wavefront gradient can be detected in
this direction).  These re-imaging optics can consist of two
cylindrical lenses as shown in Fig.~\ref{fig:design}(b).

A similar idea has previously been used for the NAOMI instrument of
the \wht \citep{2003SPIE.4839..647M}, and the SWAT \ao system
\citep{1992LLabJ...5..115B}, using a conventional lenslet array and
electronic binning in the detector (before readout), rather than
optical binning.  This will achieve a similar effect though will
suffer from higher dark current and a lower readout rate than the
\obshs described here.

\section{Practical design of an optically binned Shack-Hartmann sensor}
\label{sect:design}
When designing an \obshs, the impact of mis-alignment between the two
orthogonal sensor directions needs to be considered.  To overcome this
effect, a $90^{\circ}$ beam rotation can be introduced to the
reflected arm of the sensor, using two fold mirrors, the first of
which directs light out of the beam splitter plane, and the second
then directs light so that it is going in the same direction as the
beam transmitted through the beam splitter.  Two fold mirrors can also
be used with transmitted beam so that both beams are then at the same
height (see Fig.~\ref{fig:design}(a)), and if these fold by more than
$90^\circ$ back on themselves, the path length of the two beams can be
equalised.  Accurate alignment of these beams should be possible using
standard techniques for mirror alignment.  Both beams can then passed
through the same cylindrical lenses and lenslet array, and detected
using the same detector, which greatly simplifies the system, and
improves stability.  By using the same lenses and lenslet array for
each beam, we can ensure that truly orthogonal wavefront gradients are
measured.  Fig.~\ref{fig:design} shows a possible design for the
\obshs.  This design requires two cylindrical lenses to image the line
pattern onto the detector, compressed in one dimension.

It is also possible to use a conventional lenslet array rather than a
cylindrical (lenticular) array, if this is more readily available,
though this will require more additional optics.  In this case, after
the lenslet array, one conventional lens (to collimate the
sub-apertures), the beam splitter, two cylindrical lenses (to compress
in one dimension) and then a conventional lens (to image onto the
detector) are required for each beam.  The cylindrical lenses are used
to compress the image in one dimension, so that the sub-aperture
images in this dimension are only one pixel wide.

If a locally generated shuttered plane wave is included, injected at
the beam splitter, this can be used as a sub-aperture tilt reference
during calibration of the system, as proposed by \cite{Schmutz1979}.
By adding a calibrated tilt mirror to the reference path, both the
orthogonality and absolute magnitude of the sub-aperture tilts can be
calibrated.  This allows fine tuning of the optical train, and also
provides an empirical basis set for reconstruction matrix generation,
matched to the actual hardware.

\subsection{Advantages of optical binning}
There are several advantages for an \obshs related to the reduced
number of detector pixels required, in addition to those previously
mentioned \citep{cylindricalSH}.  For example, if a conventional \shs
uses $8\times 8$ pixels per sub-aperture, the equivalent optically
binned sensor will require eight pixels per sub-aperture for two
orthogonal directions (16 pixels in total), or fewer if advanced line
detection algorithms are used.  This means that smaller detector
arrays can be used, and frame rate can be correspondingly higher,
reducing latency due to readout time in closed loop systems, which
will improve the performance of these systems.  Additionally, the use
of fewer pixels means that detector read-noise is reduced and so the
\snr can be increased.  Computational and data bandwidth requirements
are also reduced, an important consideration for next generation
astronomical \ao systems.

Systems that may have performance improved by using an \obshs will
have $n_l>2$.  This will include any open-loop \wfs system, such as
\dm figure sensor detectors for \moao applications: Open-loop control
of deformable mirror elements by the \ao loop means that precise
knowledge of their figure at any given time is necessary, and can be
obtained using a figure sensor.  In \S\ref{sect:figuresensor}, we
discuss the requirements for one particular figure sensor.

\subsection{Disadvantages of optical binning}
The light used in each orthogonal centroid calculation is halved by
the beam splitter.  In most situations this disadvantage is overcome
by the reduction in detector readout noise due to the use of fewer
detector pixels.  However, for closed loop systems, this may not be
the case.  For such systems (once the loop has been closed), the
wavefront can be assumed to be nearly flat within each sub-aperture,
and so the Shack-Hartmann spot is close to the null position, and a
small number of pixels (typically $2\times 2$) can be used to estimate
the centroid location.  Signal and noise is therefore obtained from
four pixels.  In the optically binned case, half the light is used to
estimate the centroid location for each orthogonal direction, and
likewise half the pixels are used (two) for each direction resulting
in a lower \snr than in the conventional case.  It should be noted
that this is an extreme case, and most Shack-Hartmann sensors will use
more pixels per sub-aperture even when the loop is closed, giving an
advantage to the optically binned sensor.  The \obshs also uses extra
optics compared with a conventional \shs, two fold mirrors,
and possibly the two cylindrical lenses (most conventional systems
will contain the same number of lenses for re-imaging and scaling
purposes).  This therefore results in a slightly reduced throughput,
though reflectivity of these mirrors can be very high.

\section{Optically binned SHS performance}
We compare the performance of an \obshs with a conventional \shs using
simple analytical considerations, and Monte-Carlo simulation of a
closed loop astronomical \ao system, the results of which are
described here.

\subsection{Analytical performance estimates}
In general, the signal $s$ for the conventional \shs will be offset by
the detector readout noise and photon shot noise scaling as
$\sqrt{s+(N^2\sigma)^2}$, $N$ being the number of pixels in one
dimension of each sub-aperture and $\sigma$ being the detector readout
noise for a single pixel, giving a \snr of
$\frac{s}{\sqrt{s+(N^2\sigma)^2}}$.  For the \obshs, the signal will
be $\frac{s}{2}$, and the noise will scale as
$\sqrt{\frac{s}{2}+(N\sigma)^2}$, giving a \snr of
$\frac{s}{2\sqrt{s/2+(N\sigma)^2}}$.  The optically binned case
therefore gives better performance for $N>2$ when detector readout
noise is greater than about five electrons, for all light levels.
When detector readout noise is one electron, the \obshs gives better
performance for $N>5$ for all light levels, or $N>2$ when $s<40$
photons.  For a noiseless detector, the \obshs always gives worse \snr
performance (by a factor of $\frac{1}{\sqrt{2}}$).  For a low noise
detector with $\sigma=0.1$ electrons (e.g.\ an \emccd), the \obshs
gives better \snr performance when $N>6$ for $s>20$, which will be the
case in most practical situations.

It should be noted that two-dimensional weighted centroid calculations
\citep{2006MNRAS.371..323T} which raise the detector signal from each
pixel by a given power (typically 1.5) before spot centroid location
determination, cannot be used with the \obshs since binning of the
signal has already occurred.  However, a similar one-dimensional
weighting algorithm can be used.

\subsection{Closed loop adaptive optics Monte-Carlo simulation comparisons} 
\begin{figure}
\includegraphics[width=8cm]{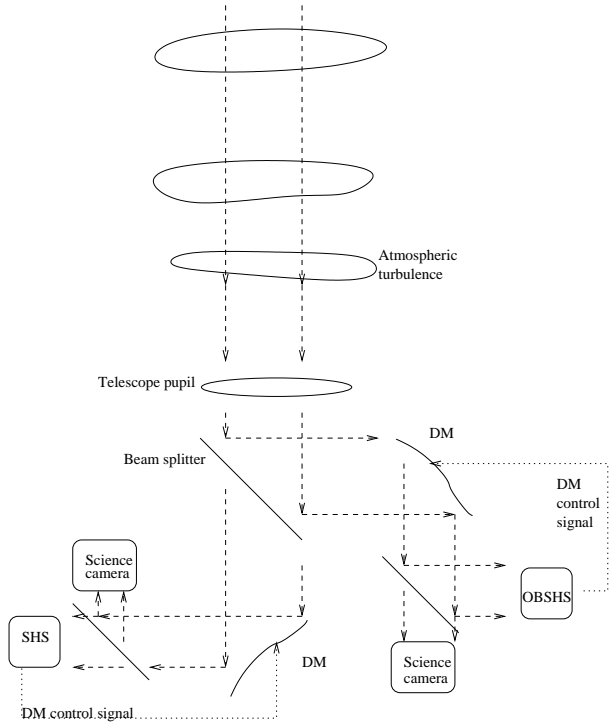}
\caption{A schematic diagram of the components of a Monte-Carlo
  simulation used to compare wavefront sensor performance.  A
  simulated beam-splitter is used to direct half the light to a
  conventional AO and imaging system, while the other half is directed
  to an optically binned AO system.  The performance of these systems
  can then be directly compared.}
\label{fig:aosim}
\end{figure}

The performance of an \obshs and a conventional \shs have been
compared using the Durham \ao simulation platform \citep{basden5}.
This software simulation comprises a classical \ao system on a 42~m
telescope for each \wfs (one \dm, one \ngs, also the science target),
and light for each \wfs passes through the same atmospheric
turbulence, as shown in Fig.~\ref{fig:aosim}.  The atmospheric
turbulence was simulated using the frozen turbulence model
\citep{taylor}, with a ground layer, and layers at 200~m and 2~km (all
uncorrelated, moving in different directions at different speeds).
Except when otherwise stated, the wavefront sensor was assumed to have
three electrons read-noise, using a 13th magnitude guide star (5~ms
integration time), $32\times 32$ sub-apertures with 8 pixels per
sub-aperture.  The effect of feedback-loop latency was not
investigated, though since this will be reduced for the \obshs (fewer
pixels), the true performance of the \obshs is likely to be further
improved relative to a conventional \shs in closed loop
situations. Further results of the simulation are not given here as
they would add length but not value to this paper, as only the
relative performance of the two \wfss are of interest here.

\subsection{Simulation results}
\begin{figure}
\includegraphics[width=8cm]{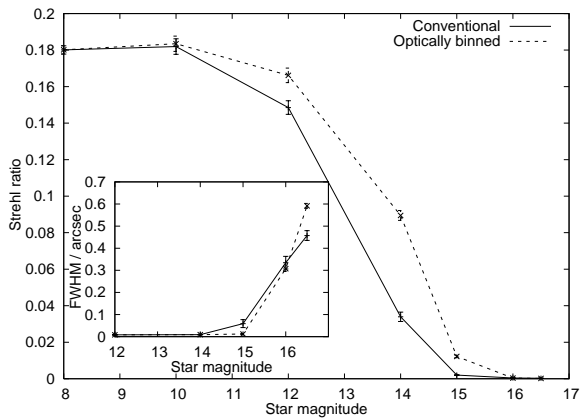}
\caption{A figure showing the relative performance between a
  conventional and optically binned SHS as a function of source
  magnitude.  The inset shows the FWHM as a function of magnitude. }
\label{fig:mag}
\end{figure}

The effect on performance of several critical \ao system parameters
has been investigated, and the Strehl ratio of an image obtained using
the \ao corrected wavefront is used as a performance estimator (a
perfectly flat wavefront will gives a Strehl ratio of unity).  The
effect of guide star magnitude on the relative performance of the
conventional \shs and the \obshs was investigated, and results shown
in Fig.~\ref{fig:mag}.  It can be seen that the \obshs offers better
relative performance as the source grows fainter (at the lowest light
level, the loop was not able to close in either case).  This is
predicted by the simple \snr calculations.

\begin{figure}
\includegraphics[width=8cm]{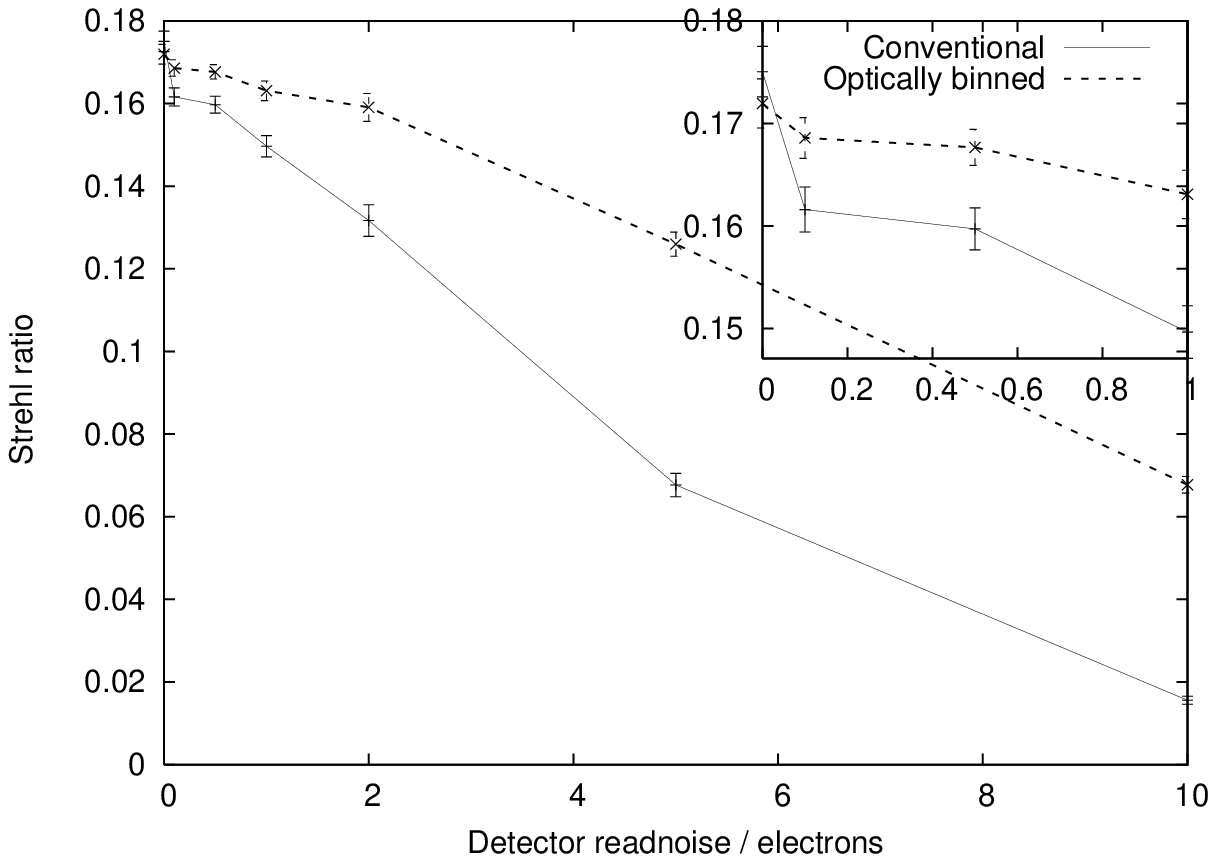}
\caption{A figure showing the relative performance between a
  conventional and optically binned SHS as a function of detector
  readout noise.  Inset is the results for 0--1 electron readout noise
  in more detail.}
\label{fig:readout}
\end{figure}

The effect of detector readout noise was also investigated, and
Fig.~\ref{fig:readout} shows that for detectors with a non-zero
readout noise, the \obshs performance is better, due to there being
fewer detector pixels required.

\begin{figure}
\includegraphics[width=8cm]{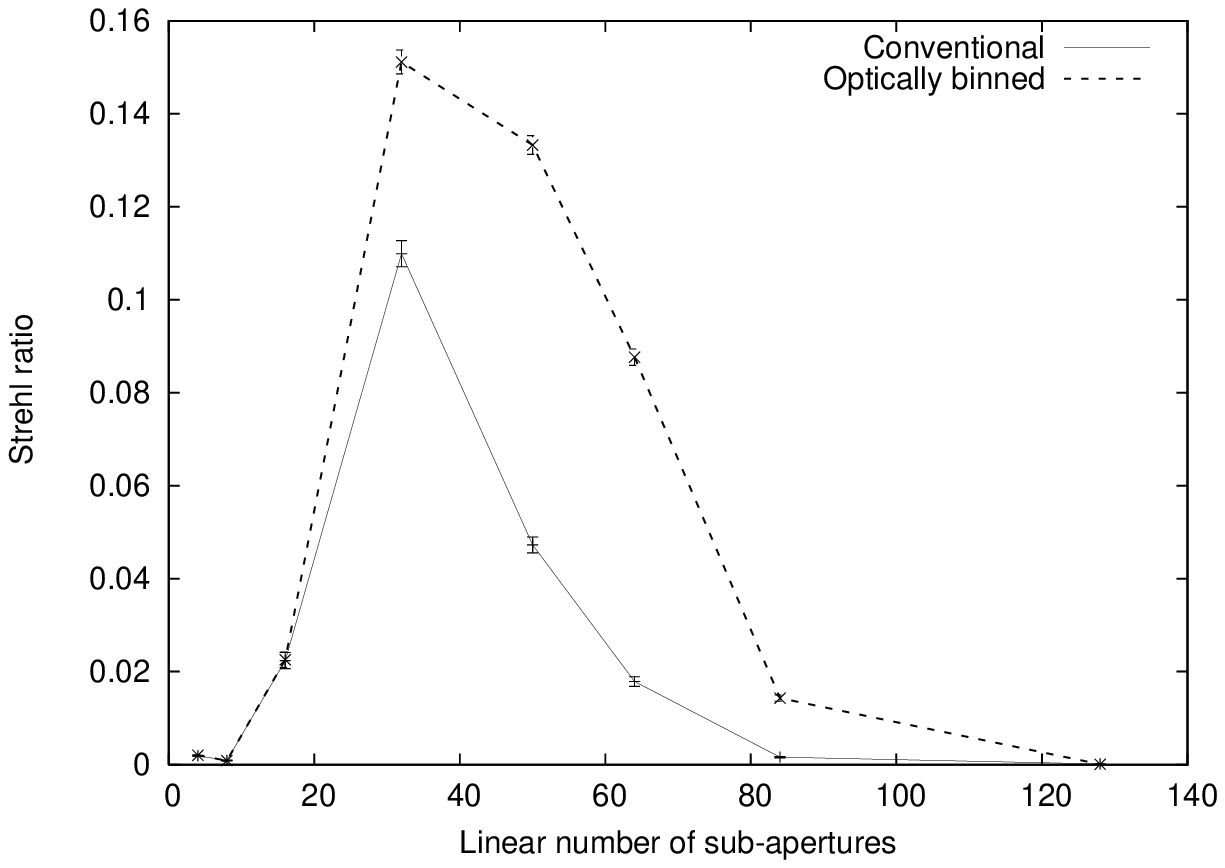}
\caption{A figure showing the relative performance between a
  conventional and optically binned SHS as a function of the linear number of
  sub-apertures.}
\label{fig:nsubx}
\end{figure}

The effect on performance of the number of sub-apertures used is shown
in Fig.~\ref{fig:nsubx}, and shows that the \obshs gives better
performance than the conventional \shs.  Higher order correction is given as the number of
sub-apertures increases.  However, light is then shared between more
sub-apertures, and so image correction eventually becomes worse.

\begin{figure}
\includegraphics[width=8cm]{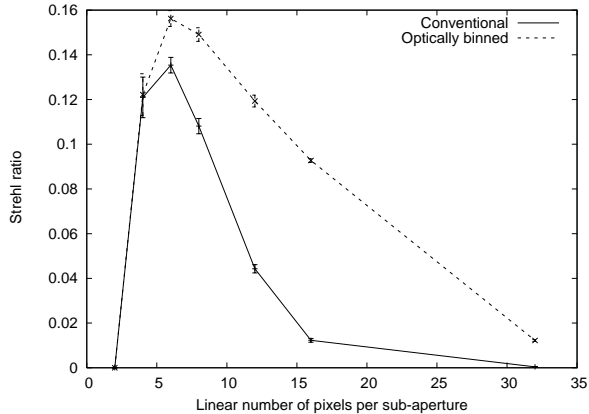}
\caption{A figure showing the dependence on relative performance
  between a conventional and optically binned SHS with the number of
  pixels per sub-aperture (linear dimension, so for the conventional
  case, the total number is this squared, while for the optically
  binned case, the total number is this multiplied by two).}
\label{fig:wfsn}
\end{figure}

Similarly, as the number of pixels per sub-aperture is varied, the
\obshs gives better performance relative to the conventional \shs, as
shown in Fig.~\ref{fig:wfsn}.  As the number of pixels increases,
light is shared between more pixels, meaning that readout noise has a
more dominant effect, resulting in poorer correction.  When too few
pixels are used, correction is also poor, as the \ao loop is difficult
to close.

\subsection{Simulation conclusions}
The extensive, but not exhaustive parameter space search carried out
to compare the performance of a conventional \shs and \obshs shows that
the \obshs can give better closed loop \ao performance in most
situations, in agreement with a simple consideration of signal and
noise sources in the \wfss.

\subsection{Case study: DUGALL figure sensor}
\label{sect:figuresensor}
The \dugall is a proposed high-order on and off-sky \lgs test
facility.  It will have several operational modes, including \moao and
\ltao.  \moao involves open loop wavefront sensing and mirror shaping
(the \wfs does not view the \dm), and relies on the shape of the
unsensed deformable mirror being known at all times.  Typically this
could involve strain gauges \citep{2004SPIE.5490...79B} or direct
measurement of the mirror shape, since hysteresis and non-linearities
within the mirror will mean that the shape of the mirror is not always
equal to the initially requested shape.  \dugall proposes to use a 4~K
actuator \mems deformable mirror, which does not contain strain
gauges, and so direct measurement of the mirror shape is required
using a figure sensor.  Figure sensing must ideally operate at several
times the rate of the \ao loop, so that several adjustments to the
mirror shape can be made during each \ao loop iteration, until the
desired shape is reached.  The \dugall \ao loop will have a maximum
operational rate of 1~kHz, and so the figure sensor may require a
frame rate of up to 10~kHz, depending on the accuracy
(time-resolution) required and control algorithms used.  Since the
figure sensor measures the shape of a non-null (non-flat) mirror,
wavefront gradients can be significant, and so a reasonable field of
view is required for each sub-aperture, giving rise to a requirement
of $8\times8$ pixels for a conventional \shs.  The 4~K deformable
mirror has $64\times64$ actuators, and so a \shs of this order is
required.  If a conventional \shs sensor is used, this would require a
detector capable of reading a $512\times512$ array at 10~kHz, and a
platform for computing centroid locations in real-time, requiring a
data rate of 5~GBs$^{-1}$, assuming two bytes per pixel.  We are
unaware of a suitable commercially available sensor able to meet these
requirements.

By using an \obshs, we would require two sensors each with
$64\times512$ pixels, and a reduced data rate of about 500~MBs$^{-1}$
for each sensor.  There are several readily available commercial
sensors that would meet this requirement.  The reduced data rate makes
centroid computation less demanding in real-time, and the \obshs will
also give an improvement in \snr.  We are currently developing such a
\wfs.

\ignore{
\section{To investigate}
Monte-carlo simulations - show whether above SNR calcs are correct.

Advanced centroid algorithms eg weighted centroids.

Investigate the improvements gained by lower latency (fewer pixels).

Setup in the lab - how does it work practically.  Alignment
sensitivities etc.
}

\section{Conclusions}
We have presented an improved design for a Shack-Hartmann \wfs, using
optical binning and a cylindrical lenslet array.  This design promises
to improve the \snr for wavefront reconstruction in most situations,
and involves splitting the wavefront for detection of the orthogonal
wavefront gradients, using an optically binned \shs.  We have
presented general Monte-Carlo simulation results which show that this
\obshs can lead to better performance than a conventional \shs, and
have presented schematic designs for such a detector.  This \wfs will
be of particular interest for open loop systems where the wavefront
gradients can be large, requiring many pixels per sub-aperture for
detection, and for systems where a high frame-rate and low latency is
important.


\end{document}